\documentclass[11pt,a4paper]{article}
\usepackage{amsmath,amssymb,booktabs}
\usepackage[margin=1in]{geometry}
\usepackage[colorlinks,linkcolor=blue,citecolor=blue,urlcolor=blue]{hyperref}
\numberwithin{equation}{section}
\newcommand{\cN}{\mathcal{N}}
\newcommand{\cF}{\mathcal{F}}
\newcommand{\cR}{\mathcal{R}}
\newcommand{\cP}{\mathcal{P}}
\newcommand{\cL}{\mathcal{L}}
\newcommand{\Nt}{\mathcal{N}^\theta}

\newcommand{\dd}{\mathrm{d}}
\begin{document}
\begin{titlepage}
\begin{center}
{\Large\bf Microscopic entropy of regular BTZ black holes\\[4pt] in 3d Lovelock gravity at all orders}\\[1.2cm]
{\large Hikmet \"Oz\c{s}ahin}\\[0.4cm]
{\it Department of Physics, Faculty of Arts and Sciences,\\ Sinop University, 57000 Sinop, T\"urkiye}\\[0.3cm]
{\tt h.ozsahin@sinop.edu.tr}\\[1.5cm]
{\bf Abstract}
\end{center}
\noindent
We give a microscopic derivation of the semi-classical entropy of the black holes of 3d Lovelock gravity
with an infinite tower of regularized Lovelock corrections, which include the regular BTZ black holes found
recently. Working in the mini-superspace formalism, we show that the reduced action and its boundary term
can be obtained in closed form at all orders in terms of the generating function of the couplings.
Assuming the ground state is described by the soliton obtained from the static black hole by a double Wick
rotation, we compute the soliton mass and reproduce the semi-classical entropy from the Cardy formula without
central charges for arbitrary couplings. We then extend the mini-superspace to stationary configurations,
compute the mass and the angular momentum of the rotating black holes obtained by a boost, and verify the
Cardy formula in the presence of rotation.
\end{titlepage}
\tableofcontents
\vspace{0.5cm}

\section{Introduction}
Providing a microscopic derivation of black hole entropy is one of the most important challenges for a
quantum theory of gravity. Strominger's derivation \cite{Strominger:1997eq} is arguably the simplest
successful example since it requires neither string theory nor supersymmetry. It relies on the observation of
Brown and Henneaux \cite{Brown:1986nw} that the asymptotic symmetries of 3d general relativity (GR) with a
negative cosmological constant are generated by two copies of the Virasoro algebra with the central charges
$c^\pm=\frac{3\ell}{2G}$, where $\ell$ is the AdS radius and $G$ is Newton's constant. The entropy then follows
from the Cardy formula for the asymptotic growth of the number of states \cite{Cardy:1986ie},
\begin{equation}
S_{\rm Cardy}=2\pi\sqrt{\frac{c^+}{6}\tilde\Delta^+}+2\pi\sqrt{\frac{c^-}{6}\tilde\Delta^-},\label{cardyc}
\end{equation}
where $\tilde\Delta^\pm=\frac12(M\ell\pm J)$ are the eigenvalues of the shifted Virasoro operators
$\tilde L_0^\pm=L_0^\pm-\frac{c^\pm}{24}$, and it reproduces the semi-classical entropy of the
Ba\~nados-Teitelboim-Zanelli (BTZ) black hole \cite{Banados:1992wn,Banados:1992gq}.

It was realized in \cite{Correa:2010hf} that this derivation does not survive in the presence of scalar hair:
for the hairy black hole of \cite{Henneaux:2002wm}, the algebra of the global charges and its central
charges remain those of GR, while the entropy as a function of the mass does not. The resolution proposed in
\cite{Correa:2010hf} is to use the Cardy formula in the form that is valid when the lowest eigenvalues
$\tilde\Delta_0^\pm$ of $\tilde L_0^\pm$ are non-vanishing (see \cite{Carlip:1998qw} for a pedagogical
derivation), which makes no reference to the central charges,
\begin{equation}
S_{\rm Cardy}=4\pi\sqrt{-\tilde\Delta_0^+\tilde\Delta^+}+4\pi\sqrt{-\tilde\Delta_0^-\tilde\Delta^-},\label{cardy0}
\end{equation}
and to assume that the ground state of the sector to which a given black hole belongs is described by the
soliton obtained from the static black hole by a double Wick rotation. With $\tilde\Delta_0^\pm=\frac12M_0\ell$,
where $M_0$ is the mass of the soliton, one has
\begin{equation}
S_{\rm Cardy}=4\pi\ell\sqrt{-M_0M}\label{cardystatic}
\end{equation}
for static black holes, and
\begin{equation}
S_{\rm Cardy}=2\pi\sqrt{-M_0\ell\,(M\ell+J)}+2\pi\sqrt{-M_0\ell\,(M\ell-J)}\label{cardyrot}
\end{equation}
in the presence of rotation. This proposal was tested successfully for different self-interactions of a
minimally coupled scalar field \cite{Correa:2011dt}, for a rotating hairy black hole \cite{Correa:2012rc},
and it has been generalized in several directions
\cite{Gonzalez:2011nz,Donnay:2015iia,Detournay:2016gao,Shaghoulian:2015dwa,BravoGaete:2015wua,BravoGaete:2015iwa,BravoGaete:2017dso,Hassaine:2019uoj}.

In \cite{Alkac:2024hvu}, the validity of the proposal was established for the first time in 3d Horndeski
theories \cite{Horndeski:1974wa,Kobayashi:2019hrl}. The theories considered there are the 3d Lovelock
gravities, which follow from the 2nd- and 3rd-order Lovelock Lagrangians
\cite{Lovelock:1971yv,Padmanabhan:2013xyr} by a regularized dimensional reduction or a conformal trick
\cite{Fernandes:2020nbq,Kobayashi:2020wqy,Lu:2020iav,Hennigar:2020lsl,Hennigar:2020fkv,Hennigar:2020drx,Alkac:2022fuc},
and admit black hole solutions analogous to the higher-dimensional ones. Since the standard methods for the
computation of the soliton mass become quite cumbersome for these theories, the whole analysis was performed
in the mini-superspace formalism, where the reduced action for a suitably general static ansatz yields both
the black hole and the soliton, together with their thermodynamics.

Recently, the regularized Lovelock invariants were obtained at all orders
\cite{Fernandes:2025eoc,Fernandes:2025fnz,Colleaux:2020wfv}. The resulting 3d theory contains an infinite tower of
corrections and, when the couplings are such that the series has a finite radius of convergence, admits
black holes which are regular everywhere, with an extremal inner horizon replacing the singularity
\cite{Fernandes:2025eoc}; this is the 3d scalar-tensor counterpart of the mechanism of \cite{Bueno:2024dgm}
(see also \cite{Bueno:2021krl,Bueno:2025jgc}). In \cite{Alkac:2026ctheorem}, it was proven that this theory
admits a holographic c-theorem at all orders and for arbitrary values of the couplings, generalizing the
result of \cite{Alkac:2022zda}, and its central charge was computed. It was also shown that all the results
depend on the couplings only through two generating functions, and that the dual theory is scale-invariant
but not conformal. As a consequence, the Cardy formula in terms of the central charge (\ref{cardyc}) is not
expected to reproduce the entropy of the black holes of the theory, and the extension of the analysis of
\cite{Alkac:2024hvu} to all orders was pointed out as a natural next step in \cite{Alkac:2026ctheorem}; it was
also posed as an open problem in \cite{Fernandes:2025eoc}. For the 4d version of the theory, the formula without
central charges was shown to work for the regular black holes with planar horizon in
\cite{Cisterna:2025vxk}. The purpose of this paper is to carry out this analysis in 3d.

Our main technical result is that the mini-superspace analysis can be performed at all orders at once.
The $n$th-order term of the reduced action depends on $n$ polynomially, and the sum over the tower closes
on the generating function $h$ of \cite{Alkac:2026ctheorem} and its first two derivatives [see (\ref{IE}) and
(\ref{dB})]. With the closed form of the boundary term at hand, we find that the soliton has the mass
\begin{equation}
M_0=-\frac{1}{8G}\left[1-\frac{h(y)}{y}\right]^{-1},\qquad y\equiv\frac{L^2}{\ell^2},
\end{equation}
and vanishing entropy for arbitrary values of the couplings, and that the Cardy formula
(\ref{cardystatic}) reproduces the semi-classical entropy. By extending the mini-superspace to stationary
configurations, which was mentioned as a possible generalization in \cite{Alkac:2024hvu}, we also compute the
global charges of the rotating black holes obtained by a boost and verify (\ref{cardyrot}). Finally, we
compare the effective central charge following from the soliton mass with the central charge of
\cite{Alkac:2026ctheorem}.

The outline of the paper is as follows: In Section \ref{sec:theory}, we introduce the theory and obtain the
reduced action and the boundary term for static configurations. Section \ref{sec:bh} is devoted to the
static black hole solution, its thermodynamics and the regular examples. In Section \ref{sec:sol}, we
construct the soliton, compute its mass, verify the Cardy formula and discuss the effective central
charge. In Section \ref{sec:rot}, we study the rotating black holes. We end with a summary and an outlook in
Section \ref{sec:sum}.

\section{3d Lovelock gravity at all orders}\label{sec:theory}
We consider the theory with the action \cite{Fernandes:2025eoc,Alkac:2026ctheorem}
\begin{equation}
I=\frac{1}{16\pi G}\int\dd^3x\sqrt{-g}\left[R-2\Lambda_0+\frac{1}{L^2}\sum_{n=2}^{\infty}c_nL^{2n}\cL^{(n)}\right],\label{action}
\end{equation}
where $\Lambda_0<0$ is the bare cosmological constant, $c_n$ are dimensionless coupling constants, $L$ is the
length scale of the tower and $\cL^{(n)}$ is the regularized $n$th-order Lovelock invariant in three
dimensions,
\begin{align}
\cL^{(n)}&=G_2^{(n)}-G_3^{(n)}\Box\phi+G_4^{(n)}R+G_{4X}^{(n)}\left[(\Box\phi)^2-(\nabla_\mu\nabla_\nu\phi)^2\right],\\
G_2^{(n)}&=-2^{n+1}(n-1)X^n,\qquad G_3^{(n)}=2^nnX^{n-1},\qquad G_4^{(n)}=-\frac{2^{n-1}n}{2n-3}X^{n-1},\nonumber
\end{align}
with $X=-\frac12\partial_\mu\phi\,\partial^\mu\phi$. We follow the conventions of \cite{Alkac:2026ctheorem}: the AdS
radius of the vacuum is denoted by $\ell$, and the couplings enter through the generating function
\begin{equation}
h(x)\equiv\sum_{n=2}^{\infty}(-1)^nc_n\,x^n,\qquad P(x)\equiv1-h'(x),\label{hdef}
\end{equation}
which satisfy $h(0)=h'(0)=0$ and $P(0)=1$. It is also convenient to define the dimensionless parameter
\begin{equation}
\lambda\equiv-\Lambda_0L^2>0,\label{lambda}
\end{equation}
which is the bare cosmological constant in units of the length scale $L$. Equivalently,
$\lambda=L^2/L_0^2$ where $L_0\equiv1/\sqrt{-\Lambda_0}$ is the AdS radius corresponding to the bare
cosmological constant. In \cite{Alkac:2024hvu}, the quadratic and the cubic theories were written with $L_0$ as
the only length scale and with dimensionless couplings multiplying $L_0^{2n-2}$. Since
$\frac{1}{L^2}c_nL^{2n}=a_nL_0^{2n-2}$ with
\begin{equation}
a_n\equiv c_n\,\lambda^{n-1},
\end{equation}
the couplings of \cite{Alkac:2024hvu} are $\alpha_2=-a_2$ and $\alpha_3=-a_3/6$, and its length scale is our $L_0$.

In order to study static field configurations, we assume a line element of the form used in
\cite{Alkac:2024hvu},
\begin{equation}
\dd s^2=-\cN^2\frac{r^2}{L^2}\cF\,\dd t^2+\frac{L^2}{r^2\cF}\dd r^2+\cR^2\dd\theta^2,\label{ansatz}
\end{equation}
where $(\cN,\cF,\cR)$ and the scalar field $\phi$ depend only on the radial coordinate $r$. Denoting the
periodicities of the Euclidean time $\tau=it$ and the angular coordinate $\theta$ by $\beta$ and $\eta$,
and performing integrations by parts until there remains no term with a derivative of $\cN$, we find that
the $n$th-order contribution to the Euclidean reduced action is proportional to $\sigma^{n-2}$ times a
quadratic polynomial in $n$, where
\begin{equation}
\xi\equiv\phi',\qquad\sigma\equiv r^2\xi^2\cF=-2L^2X .
\end{equation}
As a result, the sum over $n$ can be performed for arbitrary couplings and we obtain
\begin{align}
I_E=-\frac{\eta\beta}{16\pi GL^2}\int\dd r\,\cN\bigg\{&2\cR\Big[\lambda+h-\sigma h'\Big]-rP\Big[r\cF'\cR'+2\cF(\cR'+r\cR'')\Big]\nonumber\\
&-2r^2\cF P'\sigma'\big(\cR'-\xi\cR\big)\bigg\}+B,\label{IE}
\end{align}
where $h$, $P$ and their derivatives are evaluated at $\sigma$ and $B$ is a boundary term. Requiring that the
action has an extremum on-shell fixes its variation as
\begin{align}
\delta B=\frac{\eta\beta}{16\pi GL^2}\bigg[&rP\Big(r\cN\cF'+2\cF(\cN+r\cN')\Big)\delta\cR-2r^2\cF\cN P\,\delta\cR'-r^2\cN P\cR'\,\delta\cF\nonumber\\
&-2r^2\cN\cF P'\big(\cR'-\xi\cR\big)\,\delta\sigma\bigg],\qquad\delta\sigma=r^2\xi^2\delta\cF+2r^2\xi\cF\,\delta\xi .\label{dB}
\end{align}
When the tower is truncated at $n=2$ or $n=3$, the equations (\ref{IE}) and (\ref{dB}) reduce to the
corresponding expressions for the 3d Einstein-Gauss-Bonnet theory and the 3d cubic Lovelock gravity given
in \cite{Alkac:2024hvu} upon $\cF\to\lambda\cF$, which accounts for the use of $L_0$ instead of $L$ in the ansatz
there. Note that the combinations $1+2\alpha_2r^2\xi^2\cF-18\alpha_3r^4\xi^4\cF^2$ appearing there
are nothing but the function $P$, whose positivity ensures the uniqueness of the vacuum with a non-trivial
scalar field \cite{Alkac:2026ctheorem}.

Since the Euler-Lagrange equation of $\cN$ guarantees that there is no bulk contribution to the on-shell
action, one has
\begin{equation}
I_E\Big|_{\text{on-shell}}=B(\infty)-B(r_{\rm min}),\qquad M=\partial_\beta I_E\Big|_{\text{on-shell}},\qquad S=(\beta\partial_\beta-1)I_E\Big|_{\text{on-shell}},\label{MS}
\end{equation}
where $r_{\rm min}$ is the minimum value of the radial coordinate.

\section{Static black hole}\label{sec:bh}
\subsection{Solution and thermodynamics}
The black hole ansatz
\begin{equation}
\dd s_b^2=-N^2\frac{r^2}{L^2}f_b\,\dd t_b^2+\frac{L^2}{r^2f_b}\dd r^2+r^2\dd\theta_b^2\label{bhansatz}
\end{equation}
corresponds to the choice $\cN=N$, $\cF=f_b$, $\cR=r$. With
\begin{equation}
N=1,\qquad\phi=\log\left(\frac{r}{\ell}\right),\label{scalar}
\end{equation}
one has $\xi=\frac1r$, $\sigma=f_b$ and $\cR'-\xi\cR=0$. The Euler-Lagrange equations of $\cF$ and $\xi$ are
then satisfied identically, that of $\cN$ becomes $\big[r^2\big(\lambda-f_b+h(f_b)\big)\big]'=0$, and that of
$\cR$ follows from it. Therefore, the function $f_b$ is determined by
\begin{equation}
f_b-h(f_b)=\lambda\left[1-\left(\frac{r_+}{r}\right)^2\right],\label{Peq}
\end{equation}
which is the equation found in \cite{Fernandes:2025eoc} and generalizes the quadratic and the cubic equations
of \cite{Hennigar:2020fkv,Alkac:2022fuc}. It is the equation that determines the planar black holes of
higher-dimensional Lovelock gravity \cite{Cadoni:2016hhd,Hennigar:2017umz}. Since $h_b\equiv\frac{r^2}{L^2}f_b\to\frac{r^2}{\ell^2}$
at infinity, the value of the function $f_b$ at infinity is $y\equiv\frac{L^2}{\ell^2}$, and (\ref{Peq}) gives
\begin{equation}
y-h(y)=\lambda\qquad\Longleftrightarrow\qquad1+\Lambda_0\ell^2=\frac{h(y)}{y},\label{vac}
\end{equation}
which is the vacuum equation of \cite{Alkac:2026ctheorem} relating the AdS radius $\ell$ to the parameters of the
theory. We assume that it has a positive root $y$ such that $P(x)>0$ for $0\le x<y$, which ensures that
$f_b$ increases monotonically from $f_b(r_+)=0$ to $y$. The explicit form of $f_b$ is not needed in what
follows. Taking the derivative of (\ref{Peq}) and using $P(0)=1$, we get $f_b'(r_+)=\frac{2\lambda}{r_+}$,
and the inverse temperature is
\begin{equation}
\beta=\frac{4\pi}{h_b'(r_+)}=\frac{2\pi L^2}{\lambda r_+},\qquad\eta=2\pi .
\end{equation}
The variation of the boundary term (\ref{dB}) simplifies into
\begin{equation}
\delta B_b=-\frac{\beta r^2}{8GL^2}P(f_b)\,\delta f_b .\label{dBb}
\end{equation}
The variations of the function $f_b$ at infinity and at the horizon are
\begin{equation}
\delta f_b\Big|_\infty=-\frac{2\lambda r_+}{P(f_b)\,r^2}\delta r_+\Big|_\infty,\qquad\delta f_b\Big|_{r_+}=-f_b'(r_+)\,\delta r_+=-\frac{4\pi L^2}{\beta r_+^2}\delta r_+,
\end{equation}
where the second one follows from the conditions $h_b(r_+)=0$ and $(h_b+\delta h_b)(r_+ +\delta r_+)=0$. The
on-shell action is found to be
\begin{equation}
I_E^b\Big|_{\text{on-shell}}=B_b(\infty)-B_b(r_+)=\frac{\beta\lambda r_+^2}{8GL^2}-\frac{\pi r_+}{2G},
\end{equation}
from which the relations (\ref{MS}) give
\begin{equation}
M=\frac{\lambda r_+^2}{8GL^2}=\left[1-\frac{h(y)}{y}\right]\frac{r_+^2}{8G\ell^2},\qquad S=\frac{\pi r_+}{2G}.\label{MSbh}
\end{equation}
These are the results of \cite{Fernandes:2025eoc}, and they have the same form as those of the quadratic and the
cubic theories \cite{Alkac:2024hvu,Alkac:2023mvr}. In terms of the bare cosmological constant, the mass is
$M=-\frac{\Lambda_0r_+^2}{8G}$ and the couplings $c_n$ do not appear; they enter only through the relation
(\ref{vac}) between $\Lambda_0$ and the AdS radius. The first law $\delta M=T\delta S$ and the Smarr relation
$M=\frac12TS$ are satisfied.

\subsection{Regular black holes}
When the series (\ref{hdef}) is truncated at a finite order, the solution of (\ref{Peq}) is singular at $r=0$.
As shown in \cite{Fernandes:2025eoc}, regular solutions with $h_b=\mathcal O(r^2)$ as $r\to0$ exist when the
tower is infinite and the series has a finite radius of convergence. Then $f_b$ approaches a negative constant at
$r=0$, which is an extremal inner horizon. In Table \ref{tab}, we give the function determining $f_b$ and the
vacuum relation for the choices of the couplings leading to the regular black holes of
\cite{Fernandes:2025eoc}, for which the resummed Lagrangians were given in \cite{Fernandes:2025eoc,Alkac:2026ctheorem}.
For instance, for $c_n=1$ one has
\begin{equation}
h_b=\frac{r^2}{L^2}\,\frac{\lambda\left(1-\frac{r_+^2}{r^2}\right)}{1-\lambda\left(1-\frac{r_+^2}{r^2}\right)},\qquad\lambda=\frac{L^2}{\ell^2+L^2},
\end{equation}
which behaves as $h_b\simeq-\frac{r^2}{L^2}$ near $r=0$.
\begin{table}[t]
\centering
\begin{tabular}{lllll}
\toprule
$c_n$&$x-h(x)$&$\lambda=-\Lambda_0L^2$&$\dfrac{2G}{3\ell}\,c_{\rm eff}$&$\dfrac{2G}{3\ell}\,c_\infty$\\
\midrule
$1$&$\dfrac{x}{1+x}$&$\dfrac{y}{1+y}$&$1+y$&$1+\dfrac{y}{2(1+y)}+\dfrac32\sqrt y\arctan\sqrt y$\\[10pt]
$\dfrac{1}{n}$&$\log(1+x)$&$\log(1+y)$&$\dfrac{y}{\log(1+y)}$&$1+\sqrt y\arctan\sqrt y$\\[10pt]
$\dfrac{1-(-1)^n}{2n}$&$\tanh^{-1}x$&$\tanh^{-1}y$&$\dfrac{y}{\tanh^{-1}y}$&$1-\dfrac{\sqrt y}{2}\Big[\tanh^{-1}\sqrt y-\arctan\sqrt y\Big]$\\[10pt]
$\dfrac{\Gamma(\frac n2)\,\delta_{0,k}}{\sqrt\pi\,\Gamma(\frac{n+1}{2})}$&$\dfrac{x}{\sqrt{1-x^2}}$&$\dfrac{y}{\sqrt{1-y^2}}$&$\sqrt{1-y^2}$&${}_2F_1\big(-\frac14,\frac32;\frac34;y^2\big)$\\
\bottomrule
\end{tabular}
\caption{The function determining the black hole solution through (\ref{Peq}), the vacuum relation
(\ref{vac}) with $y=L^2/\ell^2$, the effective central charge (\ref{ceff}) and the central charge $c_\infty$ of
\cite{Alkac:2026ctheorem} for the theories admitting the regular black holes of \cite{Fernandes:2025eoc}. In the last
row, $k=(n-1)\,{\rm mod}\,2$.}\label{tab}
\end{table}

Two remarks are in order. First, the series (\ref{hdef}) converges for $|x|<1$ in these examples, whereas
$0\le f_b<y$ outside the event horizon. Therefore, the series converges in the whole exterior region only
if $\ell>L$, which is always the case in the last two rows of Table \ref{tab}. Otherwise, the theory should be
understood as defined by the resummed Lagrangian, i.e., by the analytic continuation of $h$. Our results
are not affected since (\ref{IE}) and (\ref{dB}) depend only on the function $h$. Second, the thermodynamic
analysis above and the construction of the soliton below involve only the region $r\ge r_+$. Therefore,
they hold irrespective of whether the black hole is regular, and the regular black holes are covered as
particular cases.

\section{Soliton and the Cardy formula}\label{sec:sol}
\subsection{Soliton}
Applying the double Wick rotation
\begin{equation}
t_b\to iL\theta_s,\qquad\theta_b\to i\frac{t_s}{L},\label{wick}
\end{equation}
to the ansatz (\ref{bhansatz}), one gets
\begin{equation}
\dd s_s^2=-\frac{r^2}{L^2}\dd t_s^2+\frac{L^2}{r^2f_s}\dd r^2+N^2r^2f_s\,\dd\theta_s^2,
\end{equation}
which corresponds to the following choice of the mini-superspace variables
\begin{equation}
\cN=\frac{1}{\sqrt{f_s}},\qquad\cF=f_s,\qquad\cR=Nr\sqrt{f_s}.\label{solchoice}
\end{equation}
With $N=1$ and the scalar field (\ref{scalar}), the Euler-Lagrange equations of $\cN$, $\cR$ and $\cF$
following from (\ref{IE}) become
\begin{equation}
\Big[r^2\big(f_s-h(f_s)\big)\Big]''=2\lambda,\qquad\Big[r^2\big(\lambda-f_s+h(f_s)\big)\Big]'=0,\qquad\Big[r^3P(f_s)f_s'\Big]'=0,
\end{equation}
respectively, and that of $\xi$ is satisfied identically. Hence, the soliton is described by
\begin{equation}
N=1,\qquad\phi=\log\left(\frac r\ell\right),\qquad f_s-h(f_s)=\lambda\left[1-\left(\frac{r_+}{r}\right)^2\right],
\end{equation}
where $r_+$ is now an integration constant to be fixed. Regularity of the spacetime at $r=r_+$ does not
constrain the periodicity $\beta_s$ of the Euclidean time but fixes the periodicity of the angular coordinate
as
\begin{equation}
\eta_s=\frac{4\pi L}{r_+^2f_s'(r_+)}=\frac{2\pi L}{\lambda r_+}.
\end{equation}
The boundary geometries of the black hole and the soliton are the same provided that
\begin{equation}
\beta_s=\sqrt{y}\,\beta,\qquad\eta_s=\frac{2\pi}{\sqrt{y}},\qquad\eta_s\beta_s=2\pi\beta,\label{match}
\end{equation}
which fixes the integration constant as
\begin{equation}
r_+=\frac{\sqrt{y}\,L}{\lambda}=-\frac{1}{\Lambda_0\ell}\qquad\text{for the soliton}.\label{rpsol}
\end{equation}
As in \cite{Alkac:2024hvu}, this value should be used only after integrating the boundary term.

For the choice (\ref{solchoice}), one has $\cR'-\xi\cR=\frac{rf_s'}{2\sqrt{f_s}}\neq0$ and the term with $P'$ in
(\ref{dB}) contributes, unlike the case of the black hole. Nevertheless, the four terms combine into a total
derivative and we find
\begin{equation}
\delta B_s=-\frac{\beta r^2}{8GL^2}\frac{\dd}{\dd r}\Big[rP(f_s)\,\delta f_s\Big],\label{dBs}
\end{equation}
which generalizes the expressions of \cite{Alkac:2024hvu} to arbitrary couplings. At infinity,
\begin{equation}
\delta f_s\Big|_\infty=-\frac{2\lambda r_+}{P(f_s)\,r^2}\delta r_+\Big|_\infty\qquad\Longrightarrow\qquad B_s(\infty)=-\frac{\beta\lambda r_+^2}{8GL^2}.
\end{equation}
At $r=r_+$, the variation of the function $h_s\equiv\frac{r^2}{L^2}f_s$ should respect the regularity
conditions, which imply \cite{Alkac:2024hvu}
\begin{equation}
\delta f_s\Big|_{r_+}=-f_s'(r_+)\,\delta r_+,\qquad\delta f_s'\Big|_{r_+}=-\left[f_s''(r_+)+\frac{2}{r_+}f_s'(r_+)\right]\delta r_+ .
\end{equation}
From the first two derivatives of the equation satisfied by $f_s$, we have
\begin{equation}
f_s'(r_+)=\frac{2\lambda}{r_+},\qquad f_s''(r_+)=-\frac{6\lambda+4\lambda^2P'(0)}{r_+^2},\qquad P'(0)=-2c_2 .
\end{equation}
Using them in (\ref{dBs}) gives
\begin{equation}
\Big[P\,\delta f_s+rP'f_s'\,\delta f_s+rP\,\delta f_s'\Big]_{r_+}=\Big[-2\lambda-4\lambda^2P'(0)+6\lambda+4\lambda^2P'(0)-4\lambda\Big]\frac{\delta r_+}{r_+}=0 .
\end{equation}
Only the quadratic coupling enters at this stage and its contribution cancels. Therefore, $B_s(r_+)=0$ at all
orders. Using (\ref{rpsol}), the on-shell action of the soliton is
\begin{equation}
I_E^s\Big|_{\text{on-shell}}=-\frac{\beta y}{8G\lambda},
\end{equation}
and the relations (\ref{MS}) together with (\ref{vac}) yield
\begin{equation}
M_0=-\frac{y}{8G\lambda}=-\frac{1}{8G}\left[1-\frac{h(y)}{y}\right]^{-1},\qquad S_0=0 .\label{M0}
\end{equation}
The vanishing of the entropy verifies the non-degeneracy of the ground state. For $c_n=1$, the soliton
is given by
\begin{equation}
h_s=\frac{r^2}{L^2}\,\frac{\lambda\left(1-\frac{r_+^2}{r^2}\right)}{1-\lambda\left(1-\frac{r_+^2}{r^2}\right)}\,,\qquad r_+=\frac{\ell^2+L^2}{\ell},\qquad M_0=-\frac{1}{8G}\left(1+\frac{L^2}{\ell^2}\right).
\end{equation}

\subsection{Cardy formula and the effective central charge}
Using the soliton mass (\ref{M0}) and the mass of the black hole (\ref{MSbh}) in the Cardy formula without
central charges for static black holes (\ref{cardystatic}), we find
\begin{equation}
S_{\rm Cardy}=4\pi\ell\sqrt{\frac{y}{8G\lambda}\,\frac{\lambda r_+^2}{8GL^2}}=\frac{\pi r_+}{2G},
\end{equation}
which is the semi-classical entropy in (\ref{MSbh}). The Cardy formula holds for arbitrary values of the
couplings $c_n$, as long as the vacuum equation (\ref{vac}) has a positive root with the properties stated
below it. In particular, it holds for the regular black holes of Table \ref{tab}.

The formula (\ref{cardy0}) can be written in the form (\ref{cardyc}) in terms of the effective central charge
$c_{\rm eff}=-24\tilde\Delta_0^\pm$, for which we obtain
\begin{equation}
c_{\rm eff}=-12M_0\ell=\frac{3\ell}{2G}\left[1-\frac{h(y)}{y}\right]^{-1}.\label{ceff}
\end{equation}
On the other hand, the central charge of the theory, defined as the coefficient of the logarithmic
divergence of the on-shell action on the sphere and equal to the UV value of the c-function, is
\cite{Alkac:2026ctheorem}
\begin{equation}
c_\infty=\frac{3\ell}{2G}\Big[1+g(y)\Big],\qquad g(x)\equiv\sum_{n=2}^{\infty}\frac{n(-1)^n}{2n-3}c_n\,x^{n-1}.\label{cinf}
\end{equation}
Expanding both for small $y$,
\begin{equation}
\frac{2G}{3\ell}\,c_{\rm eff}=1+\sum_{n=2}^{\infty}(-1)^nc_n\,y^{n-1}+\mathcal O(c^2),\qquad
\frac{2G}{3\ell}\,c_\infty=1+\sum_{n=2}^{\infty}\frac{n}{2n-3}(-1)^nc_n\,y^{n-1},
\end{equation}
we see that $c_\infty$ is linear in the couplings while $c_{\rm eff}$ is not, and that already the linear
terms are different except for the cubic coupling. In particular,
$\frac{2G}{3\ell}(c_{\rm eff}-c_\infty)=-c_2\,y+\mathcal O(y^2)$. Hence, the Cardy formula with the central charge
(\ref{cardyc}) does not reproduce the entropy, in agreement with the expectation of
\cite{Alkac:2026ctheorem} based on the fact that the dual theory is scale-invariant but not conformal. The two
quantities are given for the resummed theories in Table \ref{tab}. They coincide only at isolated points:
for the 3d Einstein-Gauss-Bonnet theory at $c_2\,y=\frac12$, where $P(y)=0$ and the two vacua of the theory
coincide \cite{Alkac:2024hvu}, and, for instance, at $y\simeq3.28$ for $c_n=1$ and $y\simeq28.1$ for
$c_n=\frac1n$. For the last two rows of Table \ref{tab}, where the even couplings vanish, the difference is of
order $y^4$.

We also note that the theory admits the BTZ black hole with a constant scalar field, for which all the
higher-order terms vanish. It has the AdS radius $L_0=1/\sqrt{-\Lambda_0}$, the same mass and entropy as in
(\ref{MSbh}) in terms of $r_+$, and the ground state is the global AdS$_3$ spacetime with $M_0=-\frac{1}{8G}$.
The two black holes have the same entropy as a function of the mass, but they are asymptotic to AdS
spacetimes with different radii; each sector has its own ground state, and the product of the square of
the AdS radius and the soliton mass is $\frac{1}{8G\Lambda_0}$ in both.

\section{Rotating black holes}\label{sec:rot}
\subsection{Stationary mini-superspace}
In order to include rotation, we extend the ansatz (\ref{ansatz}) as
\begin{equation}
\dd s^2=-\cN^2\frac{r^2}{L^2}\cF\,\dd t^2+\frac{L^2}{r^2\cF}\dd r^2+\cR^2\big(\dd\theta+\Nt\dd t\big)^2,\label{statansatz}
\end{equation}
where $\Nt$ also depends only on $r$. At each order, the reduced Lagrangian turns out to be independent of
$\Nt$, and quadratic in its first derivative. The sum over the tower again closes: the Lagrangian of the
static case is supplemented by the term $\frac{\cR^3P(\sigma)}{2\cN}\big(\Nt{}'\big)^2$. Introducing the
momentum conjugate to $\Nt$, the reduced action takes the Hamiltonian form
\begin{equation}
I_E=-\frac{\eta\beta}{16\pi G}\int\dd r\left[\cN\left(\mathcal H-\frac{\cP^2}{2\cR^3P}\right)-\Nt\cP'\right]+B,\qquad\cP\equiv\frac{\cR^3P}{\cN}\Nt{}',\label{IErot}
\end{equation}
where $\cN\mathcal H$ is the integrand in (\ref{IE}) together with the factor $\frac{1}{L^2}$. The Euler-Lagrange
equation of $\Nt$ is $\cP'=0$. Since the new term in the Hamiltonian constraint contains no derivatives, the
variation of the boundary term acquires only one additional term compared to (\ref{dB}),
\begin{align}
\delta B=\frac{\eta\beta}{16\pi GL^2}\bigg[&rP\Big(r\cN\cF'+2\cF(\cN+r\cN')\Big)\delta\cR-2r^2\cF\cN P\,\delta\cR'-r^2\cN P\cR'\,\delta\cF\nonumber\\
&-2r^2\cN\cF P'\big(\cR'-\xi\cR\big)\,\delta\sigma-L^2\,\Nt\,\delta\cP\bigg].\label{dBrot}
\end{align}

\subsection{Solution and thermodynamics}
Rotating solutions can be obtained from the static ones by a boost in the $t-\theta$ plane
\cite{Lemos:1994xp,Lemos:1995cm}, which we perform as
\begin{equation}
t\to\Xi\,(t-\omega\ell\theta),\qquad\theta\to\Xi\left(\theta-\omega\frac{t}{\ell}\right),\qquad\Xi\equiv\frac{1}{\sqrt{1-\omega^2}},\label{boost}
\end{equation}
with $|\omega|<1$. Applied to the black hole solution of Section \ref{sec:bh}, it gives the rotating black
holes of \cite{Fernandes:2025eoc}, which correspond to the following choice of the mini-superspace variables
\begin{equation}
\cF=f_b,\qquad\cR=\rho\equiv r\sqrt{1+\Xi^2\omega^2u},\qquad\cN=\frac{r}{\rho},\qquad\Nt=-\frac{\Xi^2\omega}{\ell}\,\frac{u}{1+\Xi^2\omega^2u},\qquad u\equiv1-\frac{f_b}{y},
\end{equation}
together with the scalar field (\ref{scalar}), where $f_b$ still satisfies (\ref{Peq}). We have checked that
this configuration solves all the Euler-Lagrange equations following from (\ref{IErot}) for an arbitrary
function $h$, and that
\begin{equation}
\cP=\frac{2\Xi^2\omega\,\ell\,\lambda\,r_+^2}{L^2}
\end{equation}
is constant. Note that the use of the AdS radius $\ell$ in the boost (\ref{boost}) ensures that $\cN\to1$ and
$\Nt\to0$ at infinity.

The event horizon is located at $r=r_+$. Avoiding a conical singularity in the Euclidean section gives
$\beta=\frac{4\pi}{\cN h_b'}\big|_{r_+}$, and the angular velocity of the horizon is $\Omega=-\Nt(r_+)$, i.e.,
\begin{equation}
T=\frac{\lambda r_+}{2\pi\Xi L^2},\qquad\Omega=\frac{\omega}{\ell}.
\end{equation}
At infinity, we have
\begin{equation}
f_b\simeq y-\frac{a}{r^2},\qquad\rho\simeq r+\frac{b}{2r},\qquad a\equiv\frac{\lambda r_+^2}{P(y)},\qquad b\equiv\frac{\Xi^2\omega^2a}{y},
\end{equation}
and the first three terms in (\ref{dBrot}) contribute as $P(y)\,(y\,\delta b+y\,\delta b+\delta a)$, whereas the
term with $P'$ and the last term in (\ref{dBrot}) vanish. The factor $P(y)$ cancels and we get
\begin{equation}
\delta B(\infty)=\frac{\beta\lambda}{8GL^2}\,\delta\left[\frac{1+\omega^2}{1-\omega^2}\,r_+^2\right].
\end{equation}
At the horizon, apart from the last term, only the terms with $\delta\cR$ and $\delta\cF$ survive in (\ref{dBrot}). Using $P(0)=1$ and
$\delta\cF|_{r_+}=-\cF'(r_+)\,\delta r_+$, they combine into the variation of the horizon radius,
\begin{equation}
\frac{\eta\beta}{16\pi GL^2}\,r_+^2\,\cN\cF'\Big(\delta\cR+\cR'\delta r_+\Big)\Big|_{r_+}=\frac{\eta}{4G}\,\delta\big[\cR(r_+)\big].
\end{equation}
This holds for any $\cR$, which shows that the entropy is given by the area law for all configurations in
the stationary mini-superspace. The last term in (\ref{dBrot}) yields
$\beta\Omega\,\delta\big(\frac{\eta\cP}{16\pi G}\big)$. As a result, the on-shell action takes the form
$I_E|_{\text{on-shell}}=\beta(M-\Omega J)-S$ with
\begin{equation}
M=\frac{1+\omega^2}{1-\omega^2}\,\frac{\lambda r_+^2}{8GL^2},\qquad J=\frac{\cP}{8G}=\frac{2\omega}{1-\omega^2}\,\frac{\lambda\ell\,r_+^2}{8GL^2},\qquad S=\frac{\pi r_+}{2G\sqrt{1-\omega^2}}.\label{rotMJS}
\end{equation}
They satisfy the first law $\delta M=T\delta S+\Omega\delta J$ and the Smarr relation $M=\frac12TS+\Omega J$.
Since $|\omega|<1$, one has $|J|<M\ell$ and the extremal limit is not attained.

\subsection{Cardy formula}
Following \cite{Correa:2012rc}, we assume that the ground state is still the static soliton of Section
\ref{sec:sol}, for which $\tilde\Delta_0^\pm=\frac12M_0\ell$ with $M_0$ given in (\ref{M0}). From (\ref{rotMJS}),
\begin{equation}
M\ell\pm J=\frac{\lambda\ell\,r_+^2}{8GL^2}\,\frac{(1\pm\omega)^2}{1-\omega^2},
\end{equation}
and the Cardy formula (\ref{cardyrot}) gives
\begin{equation}
S_{\rm Cardy}=2\pi\sqrt{\frac{y\,\ell}{8G\lambda}\,\frac{\lambda\ell\,r_+^2}{8GL^2}}\;\frac{(1+\omega)+(1-\omega)}{\sqrt{1-\omega^2}}=\frac{\pi r_+}{2G\sqrt{1-\omega^2}},
\end{equation}
which is the semi-classical entropy in (\ref{rotMJS}). Equivalently, in terms of the left and right
temperatures $T_\pm=\frac{T}{1\mp\Omega\ell}$ and the effective central charge (\ref{ceff}), the entropy takes
the canonical form
\begin{equation}
S=\frac{\pi^2\ell}{3}\,c_{\rm eff}\,(T_++T_-).
\end{equation}

\section{Summary and outlook}\label{sec:sum}
We have studied the microscopic entropy of the black holes of 3d Lovelock gravity with an infinite tower of
corrections, following the proposal of \cite{Correa:2010hf} where the ground state is described by the
soliton obtained from the static black hole by a double Wick rotation. We have shown that the
mini-superspace formalism of \cite{Alkac:2024hvu} can be applied at all orders: the reduced action (\ref{IE})
and the variation of the boundary term (\ref{dB}) are given in terms of the generating function $h$ of
\cite{Alkac:2026ctheorem}. For arbitrary couplings, the soliton has the mass (\ref{M0}) and vanishing entropy,
and the Cardy formula without central charges reproduces the semi-classical entropy. This includes the
regular black holes of \cite{Fernandes:2025eoc}, and completes the step anticipated in
\cite{Fernandes:2025eoc,Alkac:2026ctheorem}. We have also extended the mini-superspace to stationary
configurations, computed the mass and the angular momentum of the rotating black holes obtained by a
boost, and verified the Cardy formula with rotation.

Together with the results of \cite{Alkac:2026ctheorem}, our analysis gives a consistent picture. The effective
central charge (\ref{ceff}) fixed by the ground state differs from the central charge (\ref{cinf}) that is
monotonic along the holographic renormalization group flows, which is what one expects for a dual theory
that is scale-invariant but not conformal. The entropy is nevertheless reproduced by the formula
(\ref{cardy0}), which makes no reference to the central charge. It would be interesting to understand which
properties of the dual theory guarantee its validity in the absence of conformal invariance.

Some comments on the scope of our results are in order. In terms of the bare cosmological constant, the
mass and the entropy of the static black hole as functions of $r_+$ are those of the BTZ black hole, and
the combination $\ell^2M_0$ appearing in the Cardy formula is independent of the couplings. Moreover, since
the analysis involves only the region outside the event horizon, the formula is not sensitive to the
resolution of the singularity. Similarly, once the formula holds for the static black hole, its validity
for the boosted solutions follows from the transformation of the global charges, which we have derived
here from the reduced action. In this sense, the non-trivial content of our work is the closed form of the
reduced action and the boundary term, the fact that the soliton mass is determined by the vacuum equation
alone at all orders, and the comparison of the resulting effective central charge with $c_\infty$.

There are several directions for future work. The rotating solutions that we have considered are obtained
by a boost and do not reach extremality; whether the theory admits rotating black holes that are not of
this form is an open question. Electric charge can be added along the lines of \cite{BravoGaete:2015iwa},
where the ground state is a magnetically charged soliton; the regular charged black holes of
\cite{Bueno:2025jgc} suggest that this is a natural problem to consider. Finally, it would be interesting to
apply the stationary mini-superspace introduced here to the 4d theory, where the static planar case was
studied in \cite{Cisterna:2025vxk}.

\section*{Statement on the use of AI tools}
\addcontentsline{toc}{section}{Statement on the use of AI tools}
In the preparation of this work, the AI assistant Claude (Anthropic) was used for the following tasks: the
symbolic computation of the reduced action and the boundary term at all orders in Sections \ref{sec:theory}
and \ref{sec:rot} with the computer algebra system SymPy, including the check of the field equations for the
black hole, soliton and rotating configurations; the numerical cross-checks of the boundary terms and of the
closed forms in Table \ref{tab}; the search of the literature; an independent check of the equations and of
the reference list of the completed draft; and the drafting and editing of the manuscript. The method, the
choice of the problem and the interpretation of the results are those of the
author(s). All the results and the text have been checked by the author(s), who take full responsibility for
the content of this paper. 

\end{document}